\documentclass{article}
\usepackage[cp1251]{inputenc}
\usepackage[english]{babel}
\usepackage{graphicx}
\usepackage{amsmath}
\usepackage{amssymb}
\usepackage{latexsym}
\usepackage{natbib}
\begin{document}
\begin{flushright}
Journal-Ref: Astronomy Letters, 2017, Vol. 43, No. 2, pp. 106-119
\end{flushright}

\begin{center}
\Large {\bf SPH simulations of structures in protoplanetary disks}\\

\vspace{1cm}
\large {\bf T.V.\,Demidova$^{1}$, V.P.\,Grinin$^{1,2}$}
\end{center}

\normalsize

1 - Pulkovo Astronomial Observatory of the Russian Academy of Sciences,
Pulkovskoe shosse 65, St. Petersburg, 196140 Russia

2 - S. Petersburg State University, V.V.\,Sobolev Astronomical Institute,
Universitetskii pr. 28, St. Petersburg, 198504 Russia\\

\begin{center}
e-mail:proxima1@list.ru
\end{center}
\normalsize
\begin{abstract}
The hydrodynamic models of the protoplanetary disk, which perturbed by the embeded low-mass companion, were calculated by our modification of GADGET-2 code. The cases of circular and eccentric orbits which can be coplanar or slightly inclined to the disk midplane were considered. The column density of test particles on the line of sight between the central star and observer was computed during the simulations. Then the column density of the circumstellar dust was calculated under the assumption that the dust and gas are well mixed with a mass ratio $1:100$. To research the influence of the disk orientation relative to the observer on the circumstellar extinction the calculations were made for four angles of inclination of the line of sight to the disk  midplane and eight directions along the azimuth. The column density in the circumstellar and circumbinary disk were computed separately. The calculations have shown the periodic variations of column density can arise both in the circumstellar and circumbinary disks. The amplitude and shape of the variation strongly depend on the parameters of the simulated system (eccentricity and inclination of the orbit, the mass ratio of the companion and star) and its orientation in space. The  results of our simulations can be used to explain the cyclic variation of the brightness of young UX Ori type stars.

\end{abstract}

\clearpage

\section{Introduction}
Many young stars are surrounded by circumstellar gas-dust disks during the evolution of which planetary systems are born. Circumstantial evidence for their existence was obtained back in the late 1980s,
before the beginning of the epoch of spaceborne and large ground-based telescopes. These were the observations of a high linear polarization in UX Ori stars during deep photometric minima that were interpreted as the influence of radiation scattered in the circumstellar disks of these stars \citep[see, e.g.,][]{1988SvAL...14..219G}. The photometric and polarimetric cycles observed in some of these stars and caused by periodic circumstellar extinction variations suggested the existence of stable large-scale structures in their protoplanetary disks \citep[see, e.g.,][]{2007ARep...51...55R}.

Direct observations of the protoplanetary disks in various spectral regions \citep[see, e.g.,][]{ 1999AJ....117.1490P, 2005ApJ...630..958G, 2005Natur.435.1067K, 2012ApJ...753...59M, 2012ApJ...760L..26M, 2012MNRAS.421.2264K} allowed their inclinations relative to the plane of the sky to be determined: some disks were observed nearly edge-on \citep[see, e.g.,][]{1995AAS...187.3205B, 1996AJ....111.1977M, 1998ApJ...502L..65S, 2000ESASP.445..133S}. A number of objects were observed at a small angle to the plane of the sky, pole-on \citep[see, e.g.,][]{2000ApJ...544..895G, 2001AJ....122.3396G, 2000ApJ...538..793K, 2003AJ....126..385C}. As the accuracy of observations increased, large-scale structures came to be identified in the disk images: spiral arms \citep{2001AJ....122.3396G, 2011ApJ...729L..17H, 2014ApJ...785L..12C, 2014ApJ...796....1T}, ring-shaped gaps \citep{1999ApJ...525L..53W, 2015ApJ...808L...3A}, bright rings \citep{2005Natur.435.1067K}, matter-free central cavities \citep{2012ApJ...753...59M, 2012ApJ...760L..26M}, vertical warps \citep{1995AAS...187.3205B, 2000ApJ...539..435H}, and density clumps \citep{1998ApJ...506L.133G}. The formation of such structures can be a consequence of the strong perturbations produced by the orbital motion of the components in young binary and multiple systems or massive planets.

The study of such processes stimulated the numerical simulations of hydrodynamic flows in the protoplanetary disks of young stars with companions.Three-dimensional (3D) hydrodynamic simulations
were performed by \citet{1996ApJ...467L..77A, 1997MNRAS.285...33B, 1997MNRAS.285..288L, 2007AstL...33..594S} using the SPH (smoothed particle hydrodynamics) method. Finite-difference schemes were used in simulating binary systems by \citet{2002A&A...387..550G, 2005ApJ...623..922O, 2010ApJ...708..485H} in the two-dimensional (2D) approximation and by \citet{2010ARep...54.1078K} in their 3D calculations. The gas flow structure in the circumstellar disk of a binary system
was described in the above papers. The circumbinary (CB) disk of the binary system is separated from the
circumstellar disks of its components by a matter-free cavity. Density waves and shocks arises at the
inner boundary of the CB disk. The spiral streams of matter (originating in the CB disk) flow toward
the circumstellar disks of the binary components, contributing to the accretion activity of the stars.
A bridge is formed between the accretion disks of the companions through which matter from the disk
of the less massive companion is transferred to the more massive one at the times of their approach at
pericenter \citep{2011Ap&SS.335..125F}.

It should be noted that the hydrodynamic processes in the protoplanetary disks of binary stars and
in the models of accretion disks around binary supermassive black holes in the nuclei of galaxies are very similar \citep[see, e.g.,][] {2008ApJ...672...83M, 2009MNRAS.393.1423C, 2013MNRAS.434.1946N, 2013MNRAS.436.2997D, 2014ApJ...783..134F, 2014PhRvD..89f4060G, 2015ApJ...807..131S, 2016ApJ...827...43M}.

Spiral waves in disks can also arise when a circumstellar gas disk interacts with a planet \citep{1998ApJ...504..983L, 2000ApJ...537L..65N, 2008A&A...486..617K, 2008MNRAS.386..973P, 2009A&A...508.1493M, 2012A&A...539A..98M, 2013A&A...556A.148P}. The formulas describing the shape of the spiral arms in such a case were proposed by \citet{2002ApJ...569..997R} and \citet{2012ApJ...748L..22M}. In models with a planet a matter-free ring is formed (near the planet's orbit) instead of the central cavity \citep{2007A&A...471.1043D}. If the planet's orbit is inclined relative to the disk plane, then a vertical warp is formed in its central parts \citep{1997MNRAS.292..896M, 1997MNRAS.285..288L, 1995MNRAS.274..987P, 2010AstL...36..808G, 2013AstL...39...26D, 2013MNRAS.431.1320X}.

$\beta$ Pic and CQ Tau are examples of the binary systems in which the inner parts of the disks are inclined relative to the periphery \citep{2004ApJ...613.1049E, 2006A&A...460..117D, 2008A&A...488..565C}. The planet itself whose orbital inclination does not coincide with the disk midplane has recently been detected around the star $\beta$ Pic \citep{2009A&A...493L..21L,  2012A&A...542A..41C}. It should be noted that the circumstellar disk of $\beta$ Pic is a debris disk that contains almost no gas. The existence of a planet in an orbit inclined to the disk in this case suggests that planets can remain for a long time in noncoplanar orbits during the evolution of the protoplanetary disk. In several cases, the perturbations by a third body located outside the disk should be taken into account to explain the observed eclipses in close binary systems surrounded by a circumbinary disk \citep[for the simulations of unusual eclipses in the young object KH15D, see, e.g.,][] {2004ApJ...603L..45W, 2004ApJ...607..913C}.

In most cases, in theoretical works the physical processes occurring in protoplanetary disks were de-
scribed and the structure of the flows and inhomogeneities were investigated. A qualitative comparison
of the theoretical calculations with the observed images was then made. Occasionally, simple schematic
models that were not based on hydrodynamic calculations were used to calculate the asymmetry in disk images \citep{2010ApJ...719.1733F}. Our works \citep{2014AstL...40..334D, 2015A&A...579A.110R}
differ in that we obtained the theoretical images of protoplanetary disks in the infrared and submillimeter spectral ranges based on the hydrodynamic models computed by the SPH method.

The perturbations in the disk produced by the motion of the companion in its orbit can also manifest themselves as circumstellar extinction variations when such systems are observed edge-on or at a small angle to the disk plane \citep{2007AstL...33..594S, 2010AstL...36..422D, 2010AstL...36..498D, 2010AstL...36..808G}. A similar study has recently been performed by \citet{2015MNRAS.454.3472T} based on a 2D model.

In our previous papers we performed 3D hydrodynamic computations using the SPH algorithm with constant softening length and time step. We used the code developed by \citet{1996Ap.....39..141S} for the investigation of interacting galaxies and modified for the simulations of protoplanetary disks \citep{2007AstL...33..594S}. During our computations we calculated the column density of circumstellar dust as a function of orbital phase at fixed azimuth and inclination of the line of sight to the disk plane. This allowed us to investigate the influence of various structures in the disk matter on the circumstellar extinction variations and light curves of UX Ori stars, i.e., young objects whose circumstellar disks are inclined at a small angle to the line of sight \citep{1991Ap&SS.186..283G, 2013A&A...551A..21K, 2016A&A...590A..96K}.

In this paper we continue to investigate the perturbations in protoplanetary disks and their influence on the circumstellar extinction using the SPH algorithm with a variable softening length and an individual time step. This allows the accuracy of hydrodynamic calculations in the inner disk regions to be increased significantly.

\begin{figure}[!h]\begin{center}
 \makebox[0.6\textwidth]{\includegraphics[scale=1]{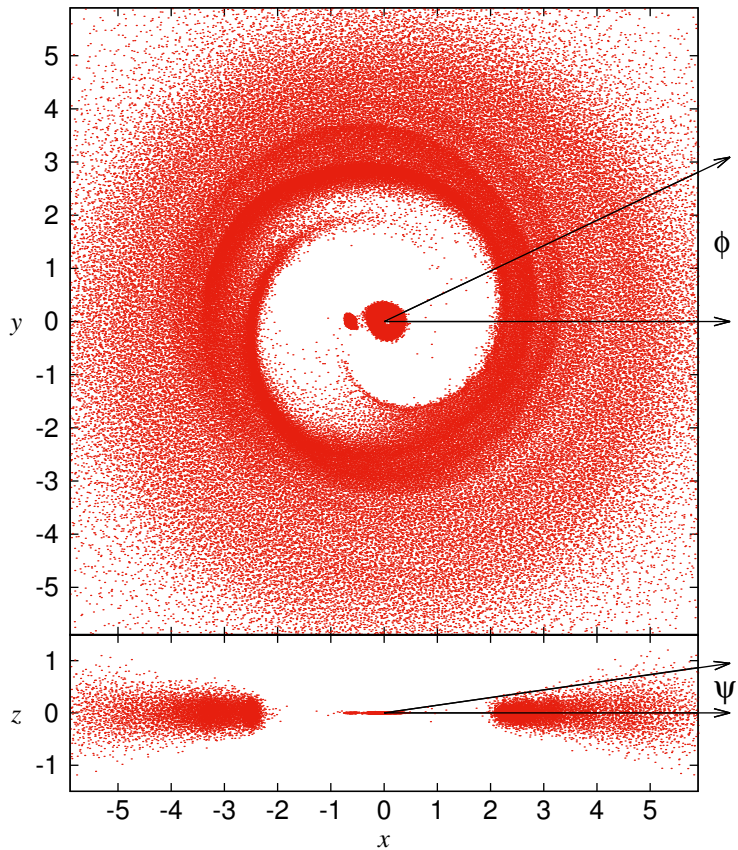}
  }
 \caption{ Distribution of matter in the disk after $20$ revolutions of the companion: a pole-on view (top) and a section along the x axis (bottom). The model: $q = 0.1$, $e = 0.3$, $\theta = 0^{\circ}$. The distances along the axes are in units of the semimajor axis of the companion's orbit.}
 \label{disk}
\end{center}
\end{figure}

\begin{figure}[!h]\begin{center}
 \makebox[0.6\textwidth]{\includegraphics[scale=1]{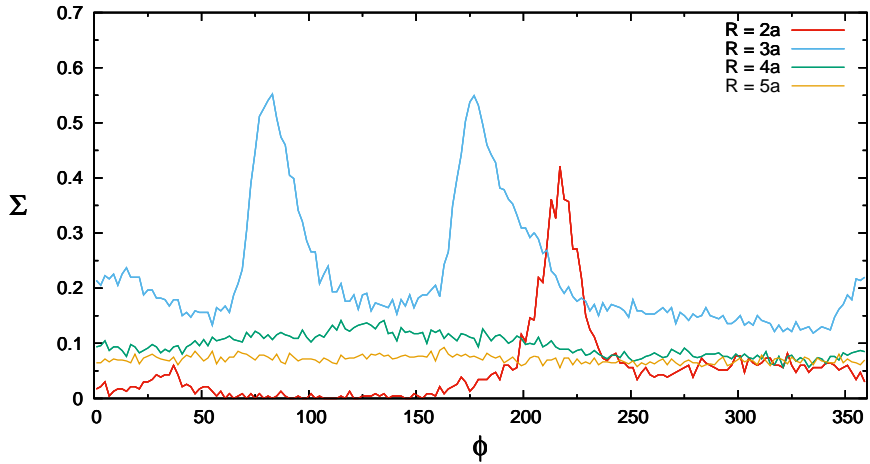}
  }
 \caption{
Surface density ($g\cdot cm^{-2}$ ) versus azimuth ($\phi$,$^\circ$) for a given radius: $R=2a$ (red line), $3a$ (blue line), $4a$ (green line), and $5a$ (orange line). The model: $q = 0.1, e = 0.3, \theta = 5^{\circ}$ after $20$ revolutions of the companion.}
 \label{sigma}
\end{center}
\end{figure}

\section{The model and method}
We consider the model of a gas-dust disk in which embedded two objects : a primary star with mass $M_1$ and a companion with mass $M_2$. A component of a binary system, a substellar companion, or a giant planet can be the companion. The main parameters of the problem are the companion-to-star mass ratio $q=M_2:M_1$, the orbital eccentricity of the companion $e$, and the orbital inclination of the companion to the disk plane $\theta$ (Fig.~\ref{disk}).

The disk mass is assumed to be small compared to the masses of the star and its companion. In this case,
the disk self-gravity exerts no strong effect on the motion of the components. Therefore, we neglected
this effect to reduce the computational time. The computations were performed in the isothermal approximation. Using this approximation is justified by the fact that the main source of periodic circumstellar extinction perturbations is a comparatively small (in extent) region of the CB disk near its inner boundary \citep{2010AstL...36..422D, 2010AstL...36..498D, 2010AstL...36..808G}. Given these assumptions, the system of hydrodynamic equations includes the continuity equation, the equation of motion, and the equation of state of an ideal gas for the isothermal case.

To numerically solve this system of equations, we use the SPH method \citep{1977AJ.....82.1013L, 1977MNRAS.181..375G}, which forms the basis for the Gadget-2 code \citep{2001NewA....6...79S, 2005MNRAS.364.1105S}. This code was developed for cosmological simulations. We modified this code and applied it to the computations of hydrodynamic flows in the models described above. A detailed description of the modified code is presented in \citet{2016Ap.....59..449D}.

By analogy with the work of \citet{1996ApJ...467L..77A}, we specified the surface density distribution at the initial time by the law $\Sigma(r) \approx r^{-1}$; the particles were distributed in height according to a barometric law with a relative disk half-thickness $\delta=z/r=0.1$. The Shakura-Sunyaev viscosity parameter was $\alpha_{ss}=0.03$. The diffusion time for the surface density evolution at the distance of the companion's semimajor axis is estimated to be $t_{\nu}\approx 1.8\cdot 10^4$ yr. The protoplanetary gas disk was simulated with $2$-$5\cdot10^5$ test particles that were placed at the initial time in a region of radius $R_d = 6 a$, where $a$ is the semimajor axis of the companion's orbit. At the same time, we considered a free outer boundary: the particles that went beyond $2R_d$ were deemed to have left the system and were excluded from our computations.

The test particle mass was specified to be $m_d=10^{25}$ g, which corresponds to an accretion rate of
$10^{-7}M_\odot yr^{-1}$, while the mass of the simulated region was $10^{-3}M_\odot$ in this case. In our previous models the test particle mass was determined as follows: the accretion rate was specified as a parameter of the problem and was compared with the accretion rate of test particles onto the binary components. The technique for calculating the accretion rate of test particles onto the star and its companion corresponds to that applied by \citet{1996ApJ...467L..77A}: in each time step the particles entering the zone whose radius $r=f\cdot R_{Roshe}$, where $f=const<1$ and $R_{Roshe}$ is the Roche lobe radius \citep[see Eq. (2) from][]{1983ApJ...268..368E}, are deemed to have been captured by
the corresponding component. The value of  $f$ in our calculations was specified to be $0.05$. To estimate the mean accretion rate in our models, the number of particles $N_i$, accreting onto the star and its companion was summed over $10$ orbital periods $P$. The relation between the accretion rate and test particle mass is then $\dot M=\frac{\sum_i N_i \cdot m_d}{10P}$. The mass of the simulated part
of the disk can also be specified as a parameter of the problem. The sound speed in the disk matter
corresponded to a temperature of $100$ K. The relaxation time of the system was assumed to be 20 orbital periods of the companion. Our computations showed that over this time interval the distribution of matter in the system comes to a stable state and ceases to depend on the initial disk geometry, a matter-free cavity is formed in the central part of the disk.

During our simulations of hydrodynamic flows we computed the accretion rate of test particles onto the
star and its companion and the column density of matter as a function of the phase of the orbital period. To estimate the influence of the disk orientation relative to the observer on the behavior of circumstellar extinction, we chose eight directions of the line of sight in azimuth (angle $\phi$) and four inclinations in elevation (angle $\psi$) (Fig. 1). Along these directions we counted the number of particles on the line of sight between the observer and central star in a column of cross section $0.1\times 0.1 a$. The column densities for the circumstellar disk of the star and the CB disk (outside the companion's orbit) were computed separately.

In Fig.~\ref{sigma} the surface density of matter at various distances $R$ from the center of mass in a $R\pm0.05a$ ring is plotted against the azimuthal angle $\phi$. Four values were considered: $R = 2a,3a,4a,5a$. The surface density is seen to be highly nonuniform in azimuth at $R = 2a$ and $3a$, but for $R = 5a$ the matter is distributed almost uniformly. Thus, a narrow layer of the disk contributes significantly to the circumstellar extinction variation, which justifies the use of the isothermal approximation for our problem. In addition, beyond $R > 5a$ the motion of the companion exerts no significant effect on the distribution of disk matter, and this allowed the computational domain to be limited to $R_d = 6a$. The orbital period of the companion was taken to be $P = 5$ yr for all models. For the adopted masses of the binary components this corresponds to the semimajor axis $a = 3.8$ AU. We disregarded the radiation from the low-mass companion when determining the photometric properties of the system.

\section{Results}
We computed a grid of models using the new code. The orbital eccentricity and inclination of the
companion were varied within the ranges $e=0-0.3$ and $\theta=0-10^{\circ}$ (the apsidal line is perpendicular to the line of nodes). The companion-to-star mass ratio for all models was chosen to be $q = 0.1$. Our computations showed that the accuracy of gas flow simulations in the central regions of protoplanetary disks with companions increased considerably compared to our previous papers \citep{2010AstL...36..422D, 2010AstL...36..498D, 2010AstL...36..808G}. Previously unresolvable flow structures became visible: the circumstellar disks of the star and its companion and the bridge between them predicted in calculations using finite-difference schemes \citep{2010ARep...54.1078K, 2010ApJ...708..485H} and identified in observations \citep{2010Sci...327..306M}.

Analysis of the computed models showed that the matter dynamics in the central parts of the system agrees well with the simulations of other authors as applied to both young binary systems and binary
black holes in galactic nuclei. It can be seen from Fig.~\ref{disk} that a cavity filled little with matter is formed in the center of the system at a distance of the order of two semimajor axes. The shape and sizes of the cavity depend on the binary component mass ratio and eccentricity. Spiral density waves arise at the inner boundary of the CB disk \citep[just as in][]{1994ApJ...421..651A, 2008ApJ...672...83M, 2013MNRAS.436.2997D}. Two streams of matter that feed the disks of the companions penetrate into the central cavity from the CB disk \citep[similar results were obtained in][]{1996ApJ...467L..77A, 2014ApJ...783..134F, 2016ApJ...827...43M}. The sizes of the accretion disks in our models also depend on eccentricity and closely correspond to the estimate from \citet{1994ApJ...421..651A}. Two-armed spiral waves (see Fig.~\ref{disks} below) described previously by other authors \citep{2000ApJ...537L..65N, 2008A&A...486..617K, 2013A&A...556A.148P, 2014ApJ...783..134F,2016ApJ...827...43M} arise and dissipate in the inner disks of the companions. A bridge exists between the accretion disks \citep{2010ARep...54.1078K, 2010ApJ...708..485H, 2014ApJ...783..134F, 2016ApJ...827...43M}.

\begin{figure}[!h]\begin{center}
 \makebox[0.6\textwidth]{\includegraphics[scale=0.55]{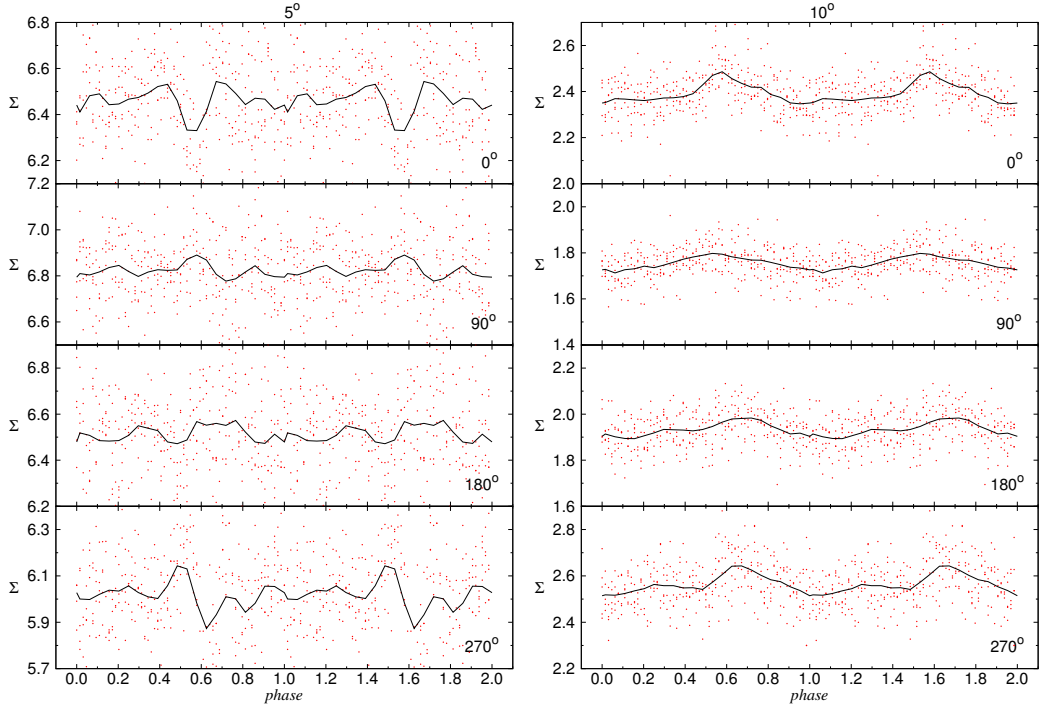}}
 \caption{Column density of matter ($g\cdot cm^{-2}$) in the CB disk of the system folded with the orbital period of the companion (dots) over five periods. The solid line indicates the averaged column density (we calculated the mean for each phase of the period with a step of $1/64P$ and then performed averaging over three points by the moving average method). The model $q = 0.1, e = 0.3,
\theta = 0^{\circ}$ . The inclination of the line of sight to the disk plane $\psi$ is $5^\circ$ (left) and $10^\circ$ (right). The azimuth $\phi$ is indicated in the lower right corner of each graph.}
 \label{col_cb}
\end{center}
\end{figure}

\begin{figure}[!h]\begin{center}
 \makebox[0.6\textwidth]{\includegraphics[scale=0.55]{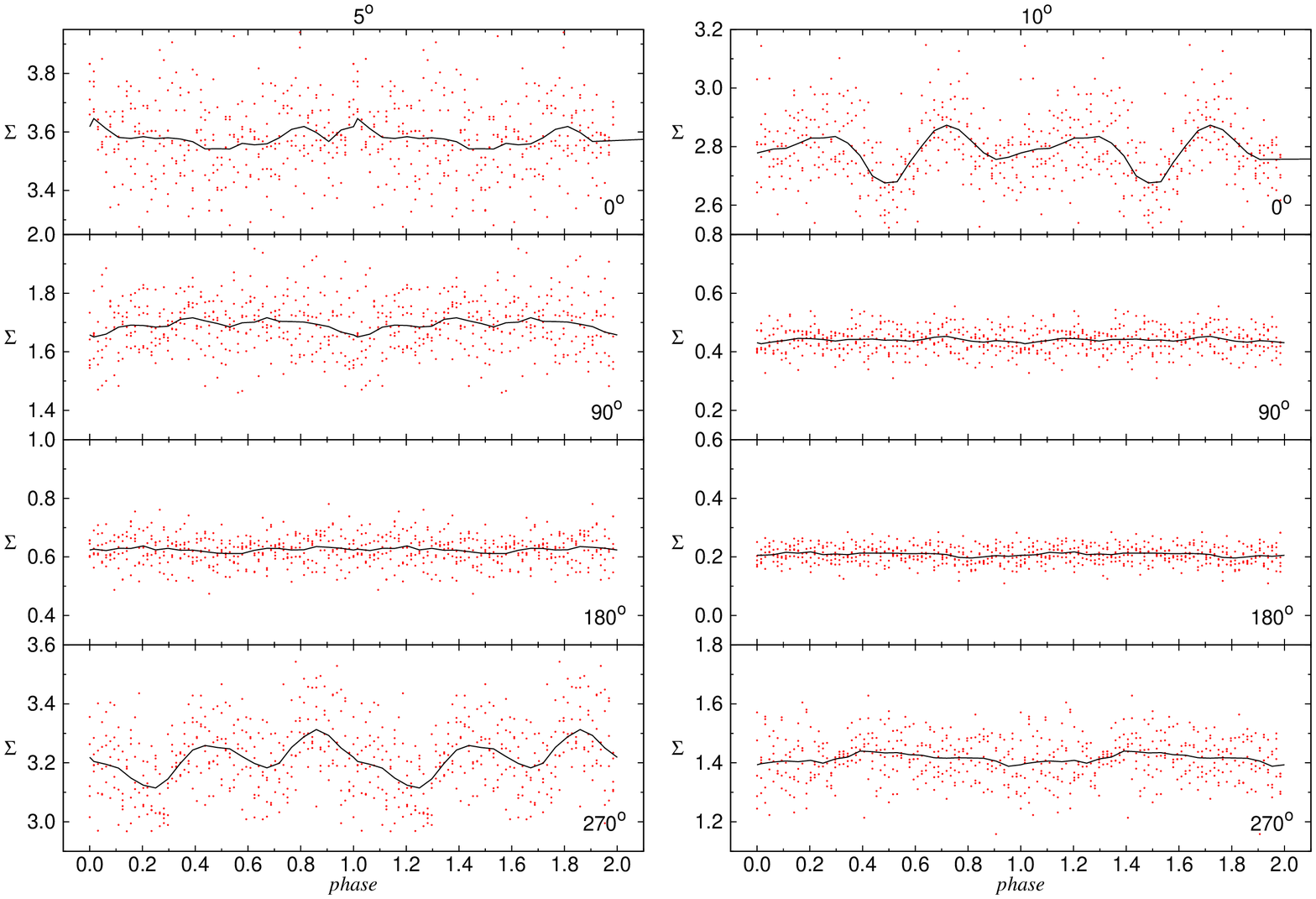}
  }
 \caption{Same as Fig.~\ref{col_cb}, for the model $q = 0.1$, $e = 0$, $\theta=5^\circ$. }
 \label{col_cbincl}
\end{center}
\end{figure}

\begin{figure}[!h]\begin{center}
 \makebox[0.6\textwidth]{\includegraphics[scale=0.55]{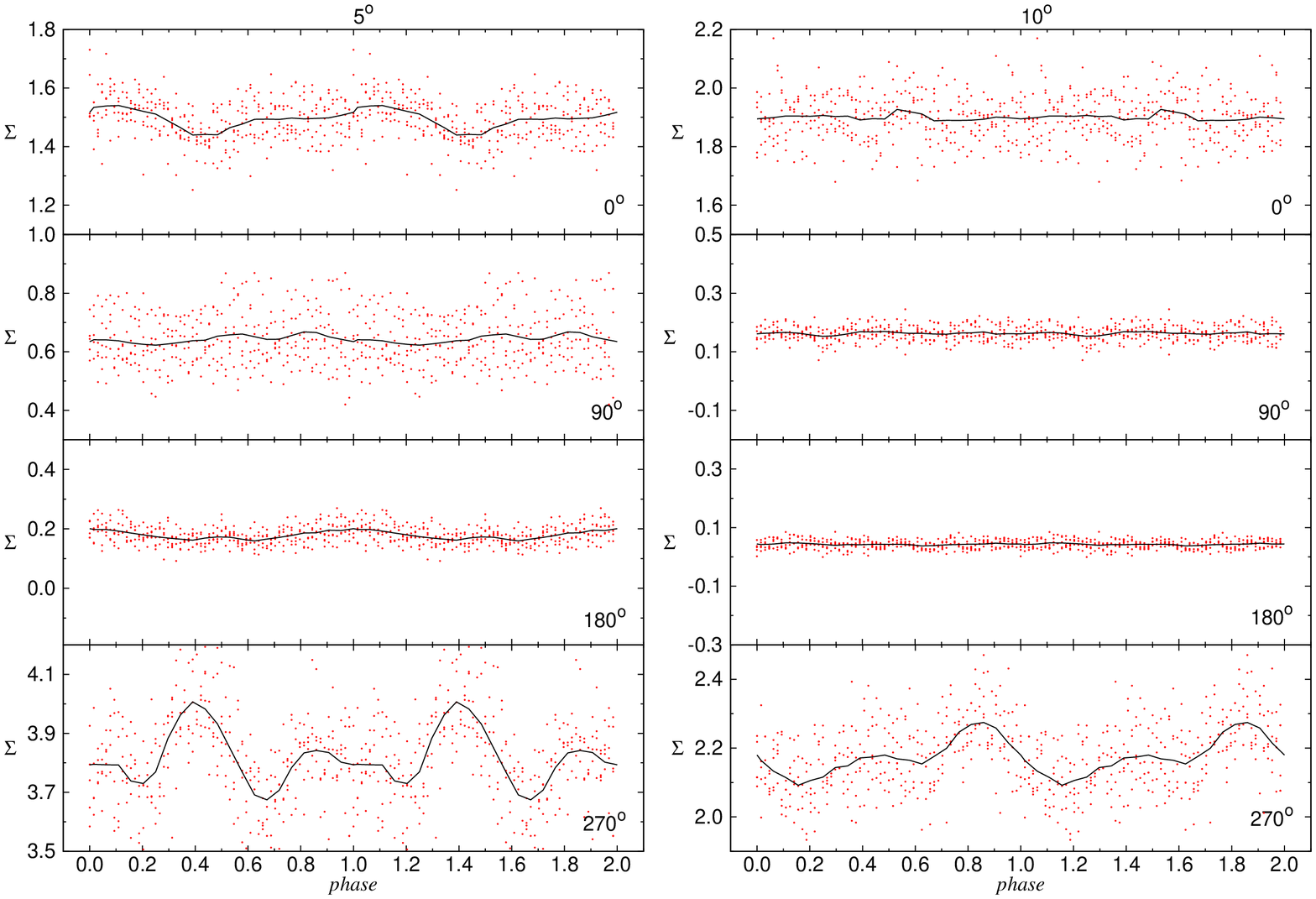}
  }
 \caption{Same as Fig.~\ref{col_cb}, for the model $q = 0.1$, $e = 0$, $\theta=10^\circ$.}
 \label{col_cbin10}
\end{center}
\end{figure}
\begin{figure}[!h]\begin{center}
 \makebox[0.6\textwidth]{\includegraphics[scale=0.7]{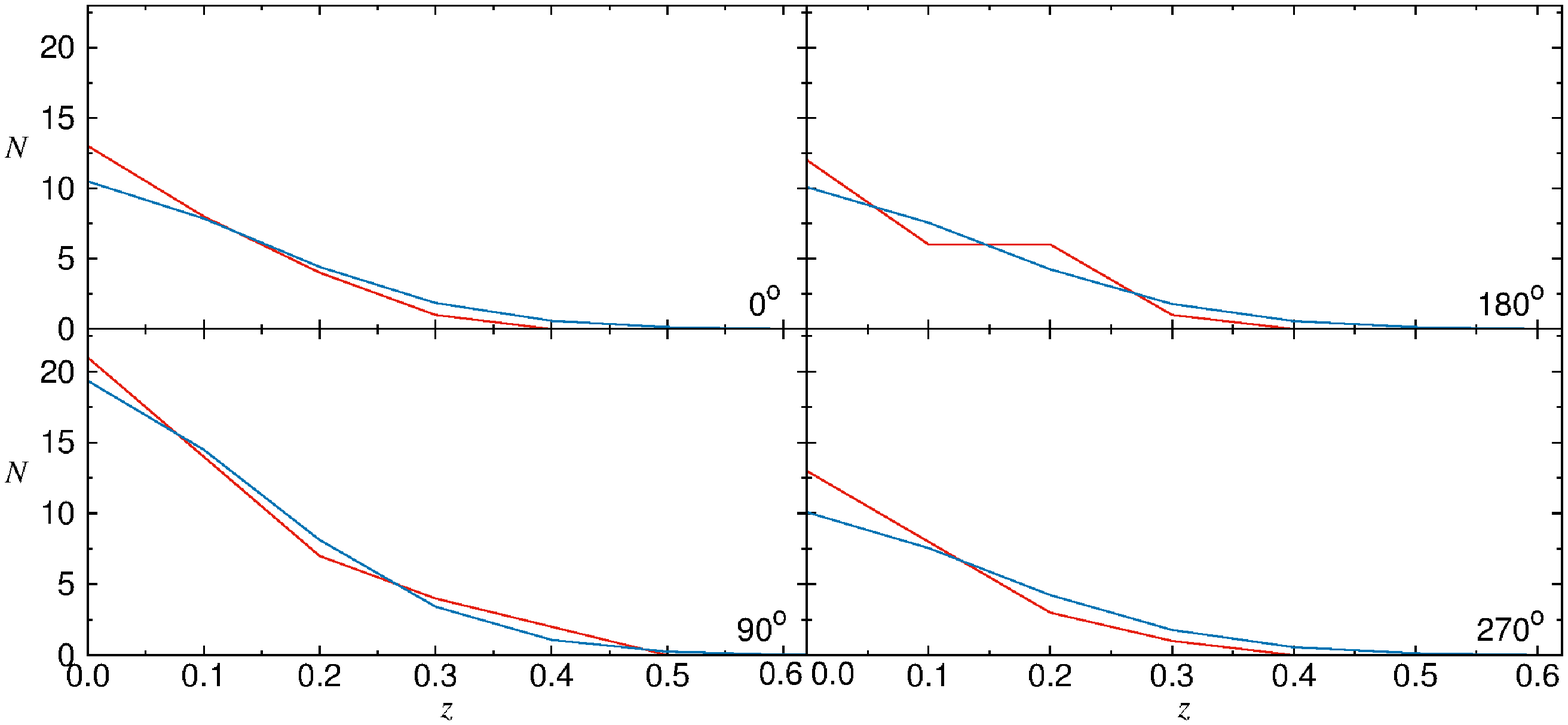}
  }
 \caption{Distributions of test particles with height above the disk plane derived from our hydrodynamic calculations (red line) and the barometric approximation (blue line) described by Eq.~(\ref{roz}). The time corresponds to $20P$, the distance from the central star is $3a$, the azimuth $\phi$ is indicated in the lower right corner of each graph.}
 \label{bar}
\end{center}
\end{figure}

\subsection{The column density in the CB disk}
Figure~\ref{col_cb} shows the behavior of the column density on the line of sight $\sigma(\phi,\psi)$ for the CB disk of a coplanar model with eccentricity $e=0.3$. Column density oscillations with orbital phase of the companion are observed for all of the considered azimuthal angles and inclinations. However, the shape of the curve depends significantly on the disk orientation relative to the observer, because density waves arise and dissipate at certain orbital phases in the case of a noncircular orbit of the companion.

The results of our computations for an inclined circular model are presented in Fig.~\ref{col_cbincl}. Periodic column density oscillations caused by the motion of the streams and density waves at the inner disk boundary are seen for some azimuthal directions. The mean column density is also seen to depend on azimuthal angle (maximally between $270^\circ$ and $0^\circ$). This is how the asymmetry of the CB disk near its inner boundary caused by the motion of the companion in an inclined orbit manifests itself the inner parts of the disk are inclined relative to its periphery. The periodic oscillations vanish for azimuthal angles of $90^\circ$ and $180^\circ$ due to the small amount of matter on the line
of sight in these directions. The maximal column density oscillations correspond to different azimuthal angles for different inclinations: $270^\circ$ and  $0^\circ$ for an inclination of $5^\circ$ and $10^\circ$, respectively. This is how the twisting of the horizontal disk layers relative to one
another manifests itself.

The amplitude of the column density oscillations increases with the orbital inclination of the companion to the disk plane (Fig.~\ref{col_cbin10}). However, at $\psi=10^\circ$ and $\psi=5^\circ$ the maximum amplitude of the column density oscillations corresponds to the same azimuthal angle of $270^\circ$. This is because in this direction the line of sight crosses the stream of matter entrained by the
companion after its passage through the highest point above the disk.

On the whole, the matter dynamics in the CB disk in the computed models agrees with our previous results \citep{2010AstL...36..422D, 2010AstL...36..498D}.

\subsection{Computing the light curves for close binaries}
A limitation of the SPH algorithm in solving the hydrodynamic equations is a small number of test
particles compared to a real gas. Therefore, the test particles have a large mass. As a result, a single
particle can contribute significantly to the circumstellar extinction, which can lead to large fluctuations when constructing the light curves. Therefore, to simulate the light curves, we passed from a discrete distribution of particles in height $z$ to a continuous density distribution described by a barometric law\footnote{A similar approach was used by \citet{2015MNRAS.454.3472T}}:
\begin{equation}
\label{roz}
\rho(x,y,z)=\rho_0(x,y)\cdot e^{-\big(\frac{z}{z_0}\big)^2}.
\end{equation}

Here, $\rho_0(x,y)$ is the density of matter in the disk plane ($z=0$), the values of $x$, $y$, and $z$ are measured from the system's center of mass. The disk half-thickness is defined by the relation $z_0=\frac{V_s}{V_\phi(r)}r$, where $V_s$ is the sound speed in the matter specified as a parameter
of the problem, and $V_\phi(r)$ is the circular Keplerian speed \citep{1973A&A....24..337S}. The sound speed was assumed to be $V_s=0.04V_\phi(a)$ (corresponding to $0.9 km\cdot s^{-1}$ and $T = 100$ K), by analogy with our previous papers and \citet{1996ApJ...467L..77A}.

It should be noted that the sound speed characterizes the disk viscosity and half-thickness. A decrease
in $V_s$ leads to a disk flattening \citep[see Fig. 1 from][]{2007AstL...33..594S} and, as a consequence, to a decrease in the optical depth on the line of sight.

Integrating Eq.~(\ref{roz}) over $z$ within the limit $z_1 = 0$ and $z_2 = \infty$, we obtain

\begin{equation}
\label{ro0}
\rho_0(x,y) = \frac{2\Sigma(x,y)}{z_0(x,y)\sqrt{\pi}}.
\end{equation}

The values of $\Sigma(x,y)$ were derived from our hydrodynamic models as follows: the test particles were summed over the height above an elementary cell of size $S = 0.1a\times0.1a$ ($a$ is the semimajor axis of the companion's orbit); the number of particles was then multiplied by the mass of a single particle and divided by the cell area $S$. Thus, all components of Eq.~(\ref{roz}) were determined, which allowed us to calculate the density of matter in the disk as a function of $x$, $y$, and $z$.

Having specified the azimuthal angle ($\phi$) and inclination ($\psi$) of the line of sight relative to the disk plane (Fig.~\ref{disk}), we calculated the density of matter $\rho_i(x,y,z)$ in the direction of $\phi$ above each cell at height $z=\sqrt{x^2+y^2}\cdot\tan\psi$. We then determined the column density on the line of sight $\sigma(\phi,\psi)$ and the optical depth in this direction $\tau(\phi,\psi)=\kappa\cdot\sigma(\phi,\psi)$, where $\kappa$ is the opacity (in our calculations we took $\kappa=200 cm^2\cdot g^{-1}$, which corresponds to Johnson's $V$ photometric band \citep[see][]{2000A&A...364..633N}. The attenuation of starlight as a result of its absorption and scattering by circumstellar dust was defined by the formula
$\Delta m(\phi,\psi)=-2.5\log e^{-\tau(\phi,\psi)}= 1.085\tau(\phi,\psi)$

To check whether it is legitimate to use the barometric approximation, we compared the distribution
of particles in height derived from our hydrodynamic computations and the barometric distribution of matter (Eq.~(\ref{roz})) at a specific radius and for a fixed time. Figure~\ref{bar} shows the height dependences of the number of particles for four azimuthal angles. Passing to the barometric approximation is seen to smooth out the particle height distribution, while the shape of the distribution is retained. It also follows from Fig.~\ref{bar} that the barometric approximation describes well the changes in the amount of matter with height at various azimuthal angles ($\phi$).

\begin{figure}[!h]\begin{center}
 \makebox[0.6\textwidth]{\includegraphics[scale=0.55]{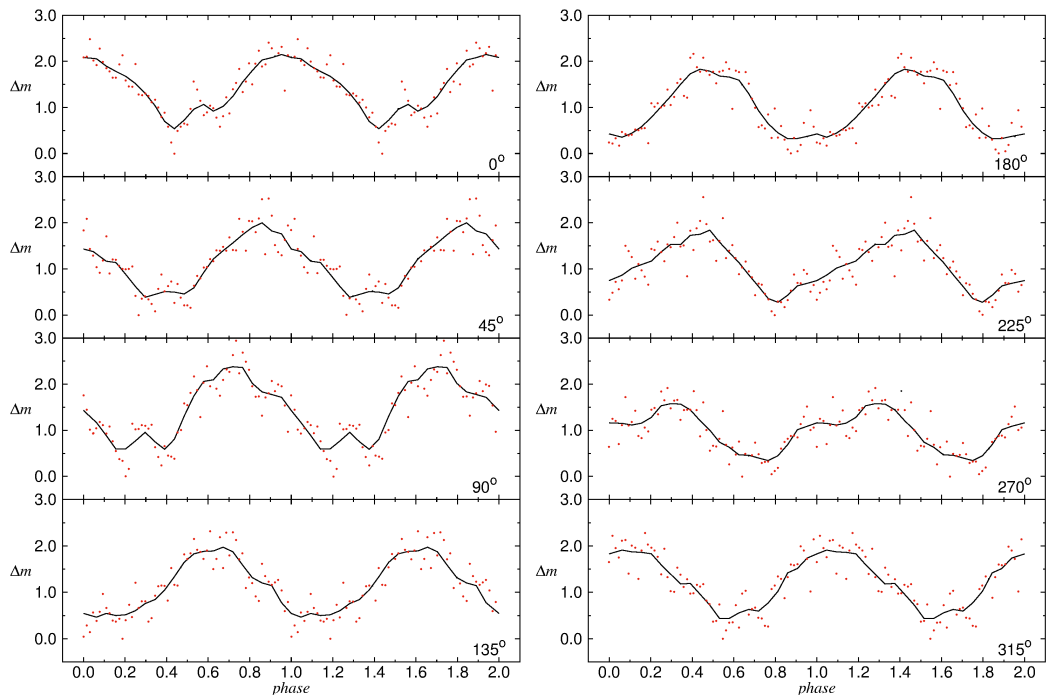}
  }
 \caption{Phase light curves due to the extinction variations in the CB disk for the model
 $q = 0.1$, $e = 0.3$, $\theta=0^\circ$, $\psi=15^\circ$.
 The azimuth $\phi$ is indicated in the lower right corner of each graph.}
 \label{brightness}
\end{center}
\end{figure}

For a coplanar model we computed the distribution of particles during one orbital period of the $P$
companion at specific times with a step of $1/64P$ and obtained $\Delta m(\phi,\psi)$ for each time step using the computational procedure described above. This allowed us to construct the phase light curves of the central star as a function of the disk orientation relative to the observer (Fig.~\ref{brightness}). Periodic brightness variations are seen on all graphs; the oscillation amplitude can reach several magnitudes even at a disk inclination of $15^\circ$ to the line of sight.
The computations described in this and previous sections are applicable to close binary systems ($a < 4$ AU) with a hot central star of spectral type $B$ or $A$, because the inner disk in this case falls into a vast dust sublimation zone and is transparent to radiation. Since the dust sublimation zone is not that large for cool T Tauri stars of spectral types from $F$ to $M$, the contribution of matter from the circumstellar disk of the primary component to the column density can be significant.

\begin{figure}[!h]\begin{center}
 \makebox[0.6\textwidth]{\includegraphics[scale=0.7]{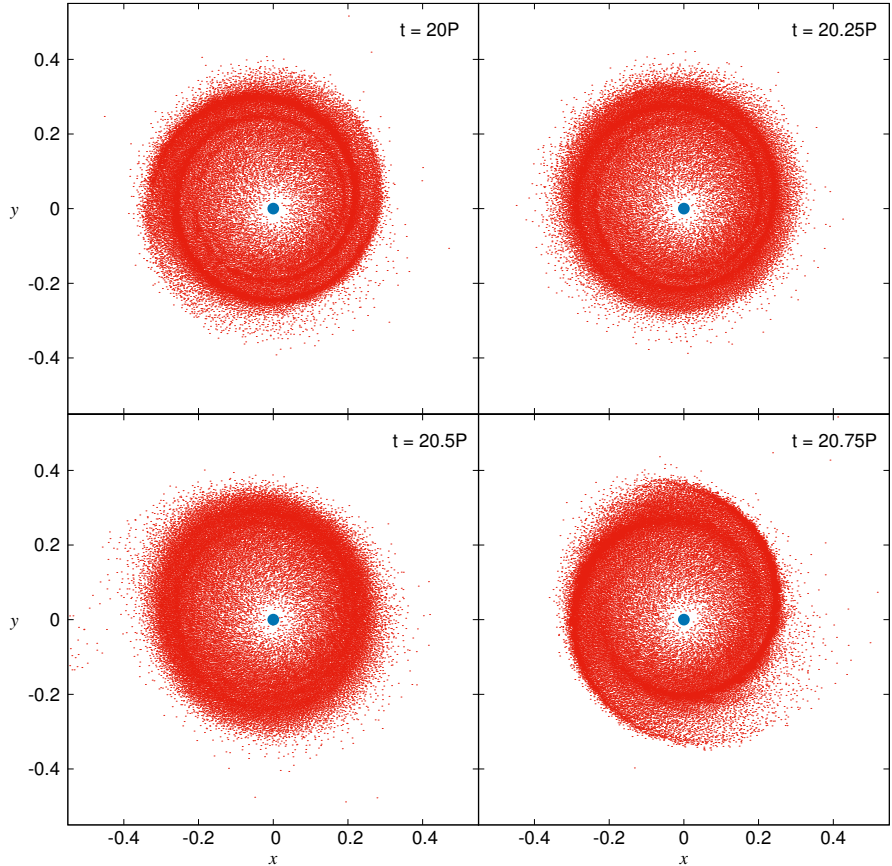}
  }
 \caption{Pole-on view of the circumstellar disk of the primary component for various orbital phases. The model: $q = 0.1$, $e = 0.3$, $\theta=5^\circ$. The time in units of the orbital period is indicated in the upper left corner: $t=20P$ corresponds to the time of apastron passage by the companion. The distances along the axes are in units of the semimajor axis of the companion's orbit.}
 \label{disks}
\end{center}
\end{figure}
\begin{figure}[!h]\begin{center}
 \makebox[0.6\textwidth]{\includegraphics[scale=0.55]{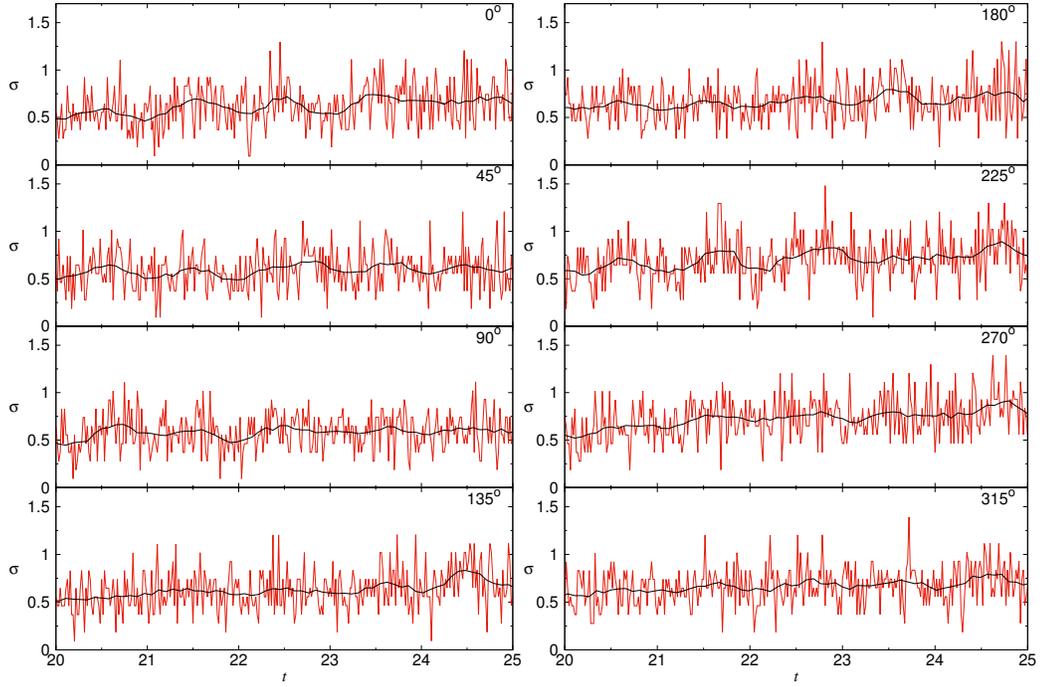}
  }
 \caption{Column density of matter ($g\cdot cm^{-2}$) in the circumstellar disk of the primary component for the model $q = 0.1$, $e = 0.3$, $\theta=0^\circ$, $\psi=10^\circ$. The azimuth $\phi$is indicated in the upper right corner of each graph. The time $t$ is given in units of the orbital period of the companion.}
 \label{col_cs}
\end{center}
\end{figure}
\begin{figure}[!h]\begin{center}
 \makebox[0.6\textwidth]{\includegraphics[scale=0.55]{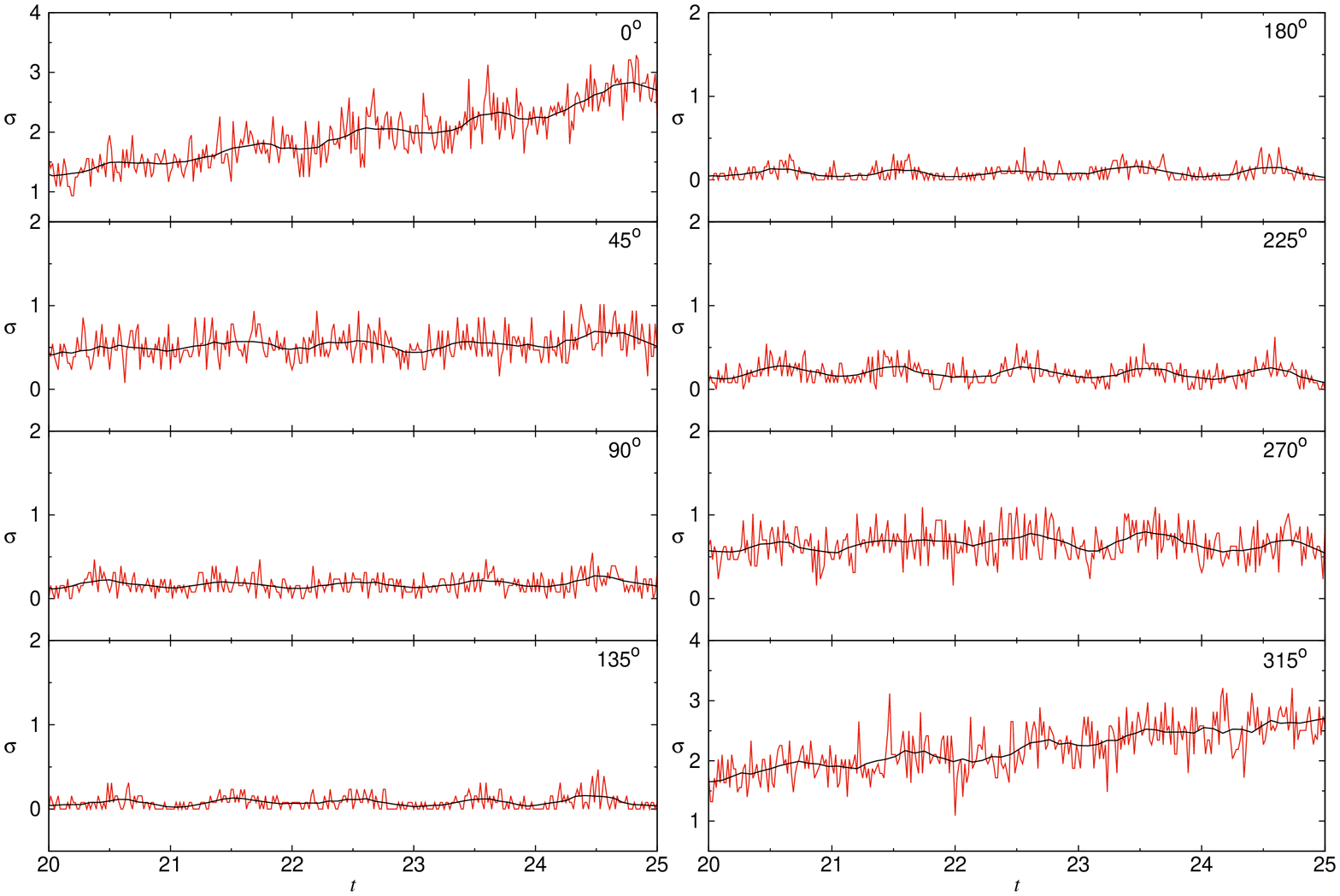}
  }
 \caption{Same as Fig.~\ref{col_csin}, for the model $q = 0.1$, $e = 0.3$, $\theta=5^\circ$, $\psi=15^\circ$. }
 \label{col_csin}
\end{center}
\end{figure}
\begin{figure}[!h]\begin{center}
 \makebox[0.6\textwidth]{\includegraphics[scale=0.8]{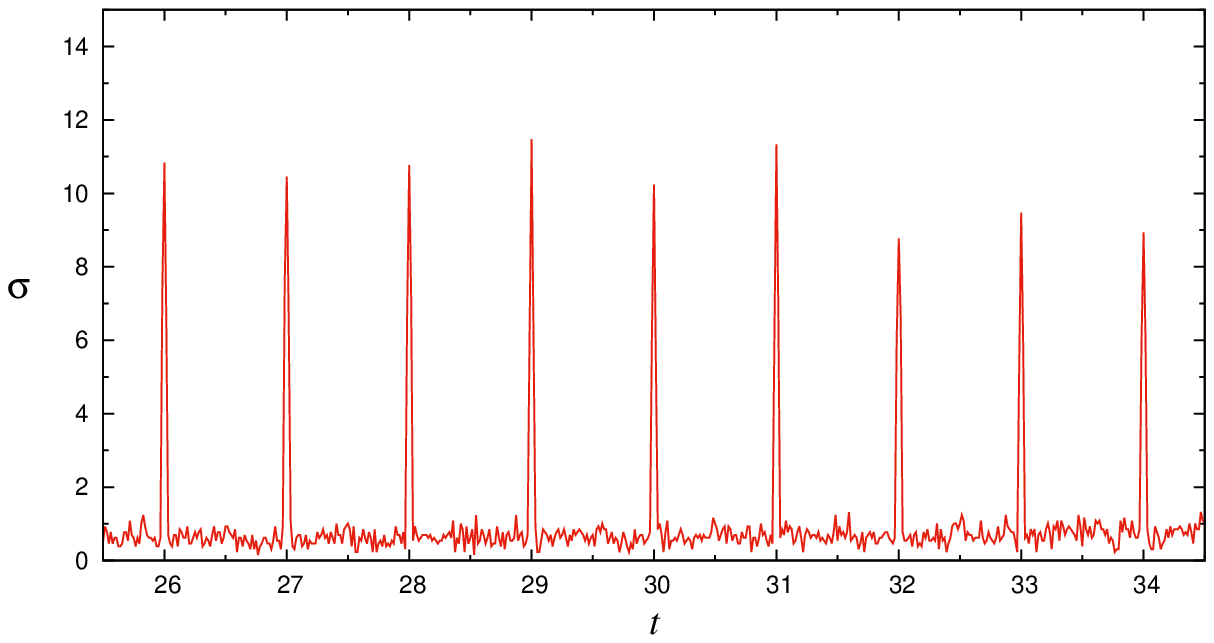}
  }
 \caption{Column density of matter ($g\cdot cm^{-2}$) in the circumstellar disk of the primary component for the model $q = 0.1$, $e = 0$, $\theta=5^\circ$, $\phi=0^\circ$, $\psi=5^\circ$. }
 \label{col_tr}
\end{center}
\end{figure}
\subsection{Extinction in the central disk regions}
A significant increase in the accuracy of calculations in the central disk region allowed us to investigate the structures emerging in the circumstellar disk of the central star. Two spiral waves produced by the tidal effect of the companion were identified. As in the papers of other authors \citep{2000ApJ...537L..65N, 2008A&A...486..617K, 2013A&A...556A.148P}, in our models these structures emerge after the periastron passage by the companion and dissipate near the apastron (Fig.~\ref{disks}).

Figure~\ref{col_cs} shows the behavior of the column density on the line of sight in the circumstellar disk for a coplanar model. The mean column density is seen to be independent of the azimuthal angle, i.e., the circumstellar disk is azimuthally symmetric. Column density oscillations with a period equal to the orbital one are noticeable in the curve. They are associated with the formation of two spiral waves in the circumstellar disk that emerge after the periastron passage by the companion and propagate toward the disk center and then dissipate after the apastron.

If we pass from the coplanar model to an inclined eccentric model, then the pattern of behavior of the
column density changes significantly (Fig.~\ref{col_csin}). A strong dependence on the azimuthal angle appears, because the circumstellar disk is inclined relative to the CB disk. A trend is traceable on the graphs corresponding to azimuthal angles of $0^\circ$ and $315^\circ$: the amount of matter on the line of sight increases noticeably with time. Such a behavior of the column density is caused by the precession of the circumstellar disk, which changes its orientation relative to the observer under the influence of the companion moving in an orbit inclined to the disk \citep[this phenomenon was first considered by][]{1997MNRAS.285..288L}. Our computations also showed that in the model significant column density oscillations could be noticeable for some azimuthal directions at inclinations of the
line of sight to the disk plane $\psi\sim15^\circ$.

In addition, a transit of the companion's disk along the line of sight can be observed at a ``lucky'' orientation of the binary system. Figure~\ref{col_tr} shows the behavior of the column density for an inclined circular model. A narrow maximum is seen to emerge in the phase dependence of the column density at the transit time of the companion's disk; its presence will give rise to a narrow minimum in the light curve of a young star.

Our computations showed that the perturbations in a comparatively narrow region near the inner boundary of the CB disk and in the circumstellar disk of the central star make a major contribution to the column density variations on the line of sight. It can be seen from the 2D models computed by \citet{2016ApJ...827...93N} for the protoplanetary disk of a binary system with components similar in mass for $2\cdot10^6$ test particles that the structure of the system's inner region differs noticeably for wide and close pairs. In the former case, streams of matter toward the circumstellar disks and a bridge between them are clearly seen. In closer pairs these structures are less distinct. The density of matter in these structures is lower than that at the inner boundary of the CB disk and in the disks of the binary components by several times. In our models the separation between the central star and its companion is considerably smaller than that in the paper of the above authors. For this reason, the bridge between the disks of the star and its companion is even less distinct (Fig.~\ref{disk})). 

\section{Conclusions}

Our computations showed that the Gadget-2 code modified by us allows the accuracy of hydrodynamic flow simulations in the central regions of the protoplanetary disks around young stars with companions
to be improved significantly. This makes it possible to investigate the influence of perturbations in the disk on the photometric properties of the models under consideration and, thus, to extend the conditions for their applicability.

In our previous papers we investigated the brightness oscillations that arose in the CB disk of the star
and its companion. In this case, the state of matter in the central part of the disk (within the companion's orbit) was calculated with a low accuracy. Therefore, the results of our computations could be used for systems in which this region fell into the dust sublimation zone and did not affect the circumstellar extinction. Herbig Ae/Be stars with companions in orbits close to the stars ($a\le4$ AU) can be attributed to such objects. The results of this paper confirm our previous computations for the CB disk and supplement them with the studies of circumstellar extinction in the central regions.

Two-armed density waves can emerge and dissipate in the circumstellar disk of the central star with
a period equal to the orbital one. This phenomenon can give rise to cyclic brightness oscillations in young stars. In addition, a transit of the companion's disk along the line of sight can be observed at a lucky orientation of the system relative to the plane of the sky. In this case, a narrow deep minimum appears in the light curve of the central star.

Large-scale inhomogeneities, both in the CB disk of a star with a companion and in the circumstellar
disk, emerge at different orbital phases for noncircular or inclined models. Therefore, to investigate the photometric properties of such systems, it is important to calculate the phase dependence of circumstellar extinction for a chosen direction rather than on the azimuthal angle.

Note that increasing the number of test particles to $2\cdot 10^6$ in the SPH simulations of a disk around a single star can lead to disk fragmentation \citep{2012MNRAS.427.2022M}. In addition, \citet{2016ApJ...827...93N} showed gas fragmentation to be possible at the inner boundary of the disk around a young binary system. In this case, the fragmentation disappears if the influence of radiation from the stars of the binary system is taken into account. To investigate the influence of such effects on the light curves of UX Ori stars, we are planning to increase the number of test particles and to pass to more self-consistent models (to take into account the disk self-gravity and thermodynamics) in future papers.

Our results can be used for more detailed simulations of cyclic brightness oscillations in young stars
whose disks are observed at a small angle to the line of sight (UX Ori stars) than those in previous papers. Our new, more accurate computations confirm that cyclic brightness oscillations in these stars can be markers of the presence of unresolved companions moving in the protoplanetary disks of these young stars.

{\bf Acknowledgments.} This work was supported by the Russian Foundation for Basic Research (project no. 15-02-09191) and the Basic Research Program 7P of the Presidium of the Russian Academy of Sciences, the  ``Experimental and Theoretical Studies of Solar System Objects and Planetary Systems of Stars'' Program
and the Program for Support of Leading Scientific Schools (NSh-7241.2016.2).
\\

\bibliography{biblio}
\end{document}